\documentstyle[aps]{revtex}
\begin{document}
\title{Exact Results for Nucleation-and-Growth in One Dimension}
\author{E.~Ben-Naim$\dag$ and P.~L.~Krapivsky$\ddag$} 
\address{$\dag$The James Franck Institute, The University of Chicago, 
Chicago, IL 60637}
\address{$\ddag$Courant Institute of Mathematical Sciences, 
New York University, New York, 10012-1185}
\maketitle
\begin{abstract}
We study statistical properties of the Kolmogorov-Avrami-Johnson-Mehl
nucleation-and-growth model in one dimension. We obtain exact results
for the gap density as well as the island distribution. When all
nucleation events occur simultaneously, the island distribution has
discontinuous derivatives on the rays $x_n(t)=nt$, $n=1,2,3\ldots$ We
introduce an accelerated growth mechanism where the velocity increases
linearly with the island size.  We solve for the inter-island gap
density and show that the system reaches complete coverage in a finite
time and that the near-critical behavior of the system is robust, {\it
i.e.}, it is insensitive to details such as the nucleation mechanism.
\end{abstract}

\section{Introduction}

Inhomogeneous systems where stable and unstable regions coexist are
common in nature. Phase separation and coarsening\cite{Bray},
aggregation \cite{Viscek}, wetting\cite{Stauffer}, dendritic
growth\cite{Pomeau}, and growth of breath figures\cite{meakin} are just
a few examples of such systems.  Typically, the stable phase grows
into the unstable phase according to complicated kinematic
rules. However, in certain cases such as adsorption \cite{Evans},
simple ballistic growth rules apply. The
Kolmogorov-Avrami-Johnson-Mehl (KAJM) nucleation-and-growth process is
a natural model incorporating nucleation of stable phases with
ballistic growth [7-15].

In this work we present exact results for various statistical
properties of the KAJM growth model in one-dimension.  The process
depends on the nucleation rate as well as the initial concentration of
the growing phase. There are two limiting cases, instantaneous
nucleation (subsequent nucleation rate vanishes) and continuous
nucleation (vanishing initial concentration). First we investigate the
KAJM growth model with a constant growth velocity. We introduce the
density of islands containing $n$ ``seeds'' and show that this
distribution is not a smooth function of the space variable. As a
result, the total island length distribution has spatial discontinuous
derivatives at every integer multiple of $t$. In the continuous
nucleation case, we obtain only the inverse Laplace transform of this
generalized island distribution. However, an asymptotic analysis shows
that the relative fraction of islands containing $n$ seeds decays
algebraically in time rather than exponentially as in the case of
instantaneous nucleation.

In the second part of our study, we introduce an accelerated
nucleation-and-growth process where the growth velocity depends on the
island size. This growth mechanism is motivated by an accelerated
random sequential adsorption (RSA) process \cite{Filipe}.  In ordinary
RSA of monomers on a lattice an adsorption attempt on a site is
successful only if that site is empty. Unlike ordinary RSA where an
attempt to adsorb on an occupied site is rejected, in the accelerated
process such an attempt is successful and adsorption takes place at
either boundary of the island. Hence, islands grow with a rate which
is a linear function of their length. The continuum version of this
model is simply the KAJM nucleation-and-growth with a growth velocity
which is linear in the island size. While for the lattice model only
an approximate theory exists, we generalize the KAJM theory to the
accelerated growth model. Exact results for the island gap
distribution show that the system is covered in a finite time. Also,
the behavior near complete coverage is robust. It is independent of
many details of the growth velocity as well as the nucleation
mechanism; it is the same for instantaneous and continuous nucleation.

The rest of this paper is organized as follows. In Section II, we
consider ordinary KAJM nucleation-and-growth process. We first review
the existing theory, and present a summary of the exact results for
both instantaneous and continuous nucleation. We then consider the
detailed island gap density and analyze its properties. In Section
III, we introduce the accelerated growth model and solve for the exact
inter-island gap distribution. Additionally, we analyze the behavior
close to complete coverage. 

\section{Theory of nucleation-and-growth in one dimension}

In the ordinary Kolmogorov-Avrami-Johnson-Mehl model size-less islands
(``seeds'') nucleate randomly in space with rate $\gamma(t)$ per unit
length, and grow with constant velocity, which we set equal to $1/2$,
in both positive and negative direction.  A collision between two such
growing islands results in a similarly growing island whose length is
given by the sum of its constituents' lengths.  Since islands grow
with unit rate, the system is covered with a rate proportional to the
density of islands, $N(t)$; in other words, the fraction of uncovered
space, $S(t)$, satisfies the rate equation
\begin{equation}
{dS(t)\over dt}=-N(t).
\label{eqst}
\end{equation}
Let us introduce $f(x,t)$, the density of inter-island gaps of size
$x$ at time $t$.  The total island density and the uncovered fraction
are simply given by $N(t)=\int_0^\infty dx\, f(x,t)$, and
$S(t)=\int_0^\infty dx\,x f(x,t)$, respectively. The gap distribution
evolves according to the following rate equation
\begin{equation}
{\partial f(x,t)\over \partial t}=
{\partial f(x,t)\over \partial x}-\gamma(t) xf(x,t)+
2\gamma(t) \int_x^\infty dy\, f(y,t).
\label{eqfxt}
\end{equation}
The first term in the right-hand side of Eq.~(\ref{eqfxt})
accounts for the shrinking of a gap caused by growth of its
two neighboring islands. The last two terms represent 
loss (gain) of gaps due to nucleation of seeds.  The rate of change
in the island density is evaluated by integrating equation (\ref{eqfxt})
\begin{equation}
{d N(t)\over dt}= -f(0,t)+\gamma(t) S(t). 
\label{eqnt}
\end{equation}
Additionally, multiplying Eq.~(\ref{eqfxt}) by $x$ and integrating, we
recover Eq.~(\ref{eqst}) thus providing a check of self-consistency.

A complementary distribution is $g(x,t)$, the density of islands of
size $x$ at time $t$.  The number density and the fraction of the
uncovered space can be alternatively expressed via the island
distribution: $N(t)=\int_0^\infty dx\, g(x,t)$ and
$S(t)=1-\int_0^\infty dx\, xg(x,t)$, respectively.  The density
$g(x,t)$ obeys\cite{Sekimoto}
\begin{equation}
{\partial g(x,t)\over \partial t}=
-{\partial g(x,t)\over \partial x}+\gamma(t) S(t)\delta(x)
+a(t)\left[\int_0^x dy\, g(y,t)g(x-y,t) -2N(t)g(x,t)\right].
\label{eqgxt}
\end{equation} 
While the first term in the right-hand side of Eq.~(\ref{eqgxt})
corresponds to growth with unit velocity, the second term represents
creation of size-less islands due to nucleation.  The last two terms
describe coalescence events between two growing islands.  Integrating
equation (\ref{eqgxt}), one finds that $N(t)=\int_0^\infty dx\,
g(x,t)$ satisfies the rate equation $dN(t)/dt=-a(t)N(t)^2+\gamma(t)
S(t)$.  Comparing this expression with Eq.~(\ref{eqnt}), the prefactor
$a(t)$ is found, $a(t)=f(0,t)/N(t)^2$.

Another interesting quantity is $g_n(x,t)$, the density of islands of
length $x$ that contain $n$ seeds. The previous island size
distribution is obtained from this more detailed quantity by a simple
summation $g(x,t)=\sum_{n\geq 1}g_n(x,t)$.  The $g_n(x,t)$
distribution obeys a generalization to Eq.~(\ref{eqgxt}):
 \begin{equation}
{\partial g_n(x,t)\over \partial t}=
-{\partial g_n(x,t)\over \partial x}+\gamma(t) S(t)\delta_{n,1}\delta(x)\
+a(t)\left[\sum_{m=1}^{n-1}\int_0^x dy\, g_m(y,t)g_{n-m}(x-y,t) 
-2N(t)g_n(x,t)\right].
\label{eqgnxt}
\end{equation} 
This equation simply reflects the fact that the number of seeds is
conserved during collisions. 

To solve the above equation, it is useful to introduce the Laplace
transform of the generating functions of $g_n(x,t)$ defined as
$g(s,z,t)=\int_0^\infty dx\, \sum_{n\geq 1} z^n e^{-sx} g_n(x,t)$.
This ``joint transform'' satisfies
\begin{equation}
{\partial g(s,z,t)\over \partial t}=
-[s+2a(t)N(t)]g(s,z,t)+z\gamma(t) S(t)+a(t)g(s,z,t)^2
\label{eqgszt}
\end{equation}
Much information concerning the island distribution can be directly
extracted from $g(s,z,t)$.  For example, the Laplace transform of the
island distribution, $g(s,t)=\int_0^\infty dx\, e^{-sx} g(x,t)$, can be
readily found from the joint Laplace transform, $g(s,t)\equiv g(s,z=1,t)$.  
The total density of islands containing $n$ seeds, $g_n(t)=\int_0^\infty dx\,
g_n(x,t)$, is obtained by considering a vanishing $s$, namely,
$g(z,t)=\sum_{n\geq 1} z^n g_n(t)=g(s=0,z,t)$.

Two natural limits of KAJM nucleation-and-growth process are
instantaneous nucleation and continuous nucleation.  In instantaneous
nucleation all seeds start growing at the same time, taken as $t=0$
for convenience.  In continuous nucleation the space contains no seeds
initially and seeds appear uniformly in space and time on yet
uncovered space. While for the former case, once the positions of the
seeds are specified the growth is fully deterministic, in the latter
case the process is stochastic.  We now present exact results for both
cases using the above formalism.

\subsection{Instantaneous Nucleation}

In instantaneous nucleation, $\gamma(t)=\Gamma \delta(t)$ since all
nucleation events occur simultaneously at $t=0$.  Hence for $t>0$,
Eq.~(\ref{eqfxt}) can be rewritten as $\partial f(x,t)/\partial x_-=0$
with $x_{\pm}\equiv x\pm t$, and the gap distribution is thus a
function of the variable $x_+=x+t$ only. Let the initial gap
distribution be $f_0(x)=f(x,t=0)$, then the solution for an arbitrary
initial distribution is readily found $f(x,t)=f_0(x+t)$.  We set
$\Gamma=1$ without loss of generality, and furthermore, we restrict our
attention to the initial conditions where the seeds are distributed
uniformly in space, $f_0(x)=e^{-x}$.  In this case, the solution
$f(x,t)=f_0(x+t)$ becomes
\begin{equation} 
f(x,t)=e^{-x-t}.  
\label{fxt1} 
\end{equation} 
The island number density as well as the uncovered fraction decay
exponentially in time $N(t)=S(t)=e^{-t}$.  The average island length
can be easily found as well $\langle x(t)\rangle=\big(1-S(t)\big)/N(t)
=e^t-1$.  These average quantities were originally derived from
simple considerations\cite{Kolmogorov,Avrami}.  For example, the
uncovered fraction equals the probability that a point, say the
origin, remains uncovered at time $t$.  For such an event to occur,
the interval $[-t/2, t/2]$ must contain no seeds initially, and hence
the $e^{-t}$ decay.  Combining $S(t)=e^{-t}$ with Eq.~(\ref{eqst})
yields the number density $N(t)=e^{-t}$.  These considerations are
applicable in arbitrary dimensions while more complete analytical
results for the gap and the island distributions are limited to one
dimension.

The Laplace transform of the joint island distribution, $g(s,z,t)$, 
satisfies 
\begin{equation}
{\partial g(s,z,t)\over \partial t}=-(s+2)g(s,z,t)+e^t g(s,z,t)^2. 
\label{eqgszt1}
\end{equation}
The above was obtained from Eq.~(\ref{eqgszt}) by substituting the 
appropriate prefactor $a(t)=f(0,t)/N(t)^2=e^t$.  
Solving Eq.~(\ref{eqgszt1}) 
subject to the initial conditions $g(s,z,t=0)=z$ gives
\begin{equation}
g(s,z,t)={ze^{-(s+2)t}\over 1-z[(1-e^{-(s+1)t})/(s+1)]}. 
\label{gszt1}
\end{equation}

The Laplace transform of the island distribution is readily found by 
evaluating $g(s,z,t)$ at $z=1$, 
\begin{equation}
g(s,t)=e^{-t}{s+1\over se^{(s+1)t}+1}. 
\label{gst1}
\end{equation}
One can immediately recover the total number density, $N(t)=g(s=0,t)=e^{-t}$,
and the fraction of uncovered space, $S(t)=1-\partial g(s,t)/\partial
s|_{s=0}=e^{-t}$.  The inverse Laplace transform can be obtained by
expanding the Laplace transform in powers of $e^{-st}$, {\it i.e.}, 
$g(s,t)=\sum_{m\geq 1} \tilde g_m(s,t)e^{-mst}$. Performing the inverse
Laplace transform term by term yields
\begin{equation}
g(x,t)=e^{-2t}\left[\delta(x-t)+\theta(x-t)\right]
+\sum_{n=1}^\infty (-1)^n e^{-(n+2)t}\theta\left(x-(n+1)t\right)
\left[{\left(x-(n+1)t\right)^{n-1}\over (n-1)!}+
{\left(x-(n+1)t\right)^n\over n!}\right],
\label{gxt1}
\end{equation}
where $\theta(x)$ is the Heavyside step function.  Unlike the gap
distribution which is a simple smooth function, $f(x,t)=e^{-x-t}$, the
island distribution is more complex.  Although it is a continuous
function, it has discontinuous derivatives on the rays $x_n(t)=nt$,
for $n=1,2,3\ldots$ These discontinuities are exponentially suppressed
because of the $e^{-nt}$ term and should be noticeable only for small
$n$ in the long time limit. This situation is reminiscent of the
behavior of extremal properties of stochastic systems such as random
walks and fragmentation models \cite{Frachebourg}.  The tail of the
distribution can be found directly from the Laplace transform. Taking
the $s=0$ limit of equation (\ref{gxt1}) and performing the inverse 
Laplace transform yields
\begin{equation}
g(x,t)\sim \langle x \rangle^{-2}e^{-x/\langle x \rangle}, \qquad x\to\infty,  
\label{gxt2}
\end{equation}
with $\langle x \rangle \sim e^t$. Hence, the tail of the island
length distribution approaches an exponential distribution with an
exponentially growing average. 
The term $e^{-2t}\delta(x-t)$ in equation (\ref{gxt1}) arises from
islands that contain a single seed. Thus, the total number density of
such islands is $g_1(t)=e^{-2t}$, or equivalently, the fraction of
one-seed islands decays exponentially in time $p_1(t)\equiv
g_1(t)/N(t)=e^{-t}$.  The rest of the $g_n(t)$ distribution is easily
obtained from the joint generating function by evaluating the $s=0$
limit, $\sum_{n\geq 1} g_n z^n=g(s=0,z,t)$ and expanding in powers of
$z^n$.  Thus we find for the fraction of $n$-seeds islands
\begin{equation}
p_n(t)\equiv {g_n(t)\over N(t)}=e^{-t}(1-e^{-t})^{n-1}. 
\label{pnt}
\end{equation}
The average number of seeds, $\langle n(t)\rangle=\sum_{n\geq 1}np_n(t)$,
is readily found to be $\langle n(t)\rangle=e^t$.  Therefore in the 
scaling limit, $t\to \infty$ with $n/\langle n(t)\rangle$ kept finite,
the exact distribution of Eq.~(\ref{pnt}) becomes
\begin{equation}
p_n(t)\simeq
\langle n(t)\rangle^{-1}e^{-n/\langle n(t)\rangle}.
\label{pnt1}
\end{equation}

To find the complete $g_n(x,t)$ distribution one first determines the
Laplace transform $g_n(s,t)=\int_0^\infty dx\,  e^{-sx}g_n(x,t)$ by expanding
$g(s,z,t)$ in powers of $z^n$,
\begin{equation}
g_n(s,t)=e^{-(s+2)t}\left({1-e^{-(s+1)t}\over s+1}\right)^{n-1}. 
\label{gnst1}
\end{equation}
Then one has to perform the inverse Laplace transform.  Simple
explicit expressions are found for small $n$, {\it e.g.}, the distribution of
``monomers'' is indeed $g_1(x,t)=e^{-2t}\delta(x-t)$ as we have
already seen previously, and the distribution of dimers is
$g_2(x,t)=e^{-(x+t)}[\theta(x-t)-\theta(x-2t)]$. Generally for $n\ge
2$, one finds
\begin{equation}
g_n(x,t)=(n-1)e^{-(x+t)}\sum_{m=0}^{n-1} {(-1)^m
[x-(m+1)t]^{n-2}\theta[x-(m+1)t]\over m!(n-1-m)!}.\qquad t\le x\le nt
\label{gnxt1}
\end{equation}
The distribution $g_n(x,t)$ vanishes outside the interval $[t, nt]$.  
An island of maximal length $nt$ results from an initial
arrangement of the $n$ seeds where all nearest neighbor distances
equal $t$.  Similar to the island distribution, $g_n(x,t)$ has
$n$ discontinuous  derivatives at  $x=t,2t,\ldots,nt$. 

It is also useful to consider the average length of an island
containing $n$ seeds, $\langle x \rangle_n=-\partial \ln
g_n(s,t)/\partial s|_{s=0}=t+(n-1)[1-t/(e^t-1)]$. As a result, the leading
behavior in both limiting cases is linear in time 
\begin{equation}
\langle x_n(t) \rangle\simeq\cases
{(n+1)t/2,  &$t\to 0$;\cr 
(n-1)+t,    &$t\to\infty$.\cr}
\label{xnt1}
\end{equation}
The prefactor depends on $n$ initially, and at the later stages of the
growth process it becomes independent of $n$. The width of the
distribution $\sigma_n^2=\langle x^2\rangle-\langle x\rangle^2$ is
conveniently obtained from Eq.~(\ref{gnst1}),
$\sigma_n^2(t)=\partial^2 \ln g_n(s,t)/\partial s^2|_{s=0}$, and we
quote only the more interesting long time asymptotics. In this limit,
the width of the distribution approaches a constant, $\sigma_n^2(t)\to
n-1$ as $t\to\infty$.

\subsection{Continuous Nucleation}  

We now turn to the case of continuous nucleation, $\gamma(t)={\rm
const}$, and set $\gamma=1$ for notational simplicity.  As a
preliminary step, we solve for the gap distribution $f(x,t)$ subject
to the initial conditions
\begin{equation}
\int_0^\infty dx\,f(x,0)=0, \qquad
\int_0^\infty dx\,x f(x,0)=1,
\label{data}
\end{equation}
corresponding to no seeds present at $t=0$.  
Substituting the ansatz $f(x,t)=\phi(t)e^{-xt}$ eliminates
the size dependence from the rate equation (\ref{eqfxt}). The time
dependent prefactor $\phi(t)$ satisfies the following ordinary
differential equation, $d\phi(t)/dt=\phi(t)(2/t-t)$.  Solving this
subject to the initial conditions of Eq.~(\ref{data}) yields
$\phi(t)=t^2e^{-t^2/2}$, and thus the gap distribution is given by
\begin{equation}
f(x,t)=t^2e^{-xt-t^2/2}. 
\label{fxt2}
\end{equation}
Integrating over the space variable gives the island number density,
the uncovered fraction, and the average island length:
\begin{equation}
N(t)=te^{-t^2/2}, \qquad
S(t)=e^{-t^2/2},  \qquad
\langle x(t)\rangle={e^{t^2/2}-1\over t}. 
\label{nsx}
\end{equation}
These average quantities can be found from simple considerations as well.
The uncovered fraction equals the probability that a point, say the
origin, remains uncovered during the time interval $[0, t]$. For this
event to happen, no nucleation events can occur at a point $x$,
$0<|x|<t/2$ during the time interval $[0, 2|x|]$.  The probability for
such an event is indeed $S(t)=e^{-t^2/2}$. The number density can also
be easily found using the relation $dS(t)/dt=-N(t)$.

Once the prefactor $a(t)=f(0,t)/N(t)^2=e^{t^2/2}$ is known the
equation satisfied by the joint transform of the gap distribution
$g(s,z,t)$ can be written
\begin{equation}
{\partial g(s,z,t)\over \partial t}=
-[s+2t]g(s,z,t)+ze^{-t^2/2}+e^{t^2/2}g(s,z,t)^2. 
\label{eqgszt2}
\end{equation}
The transformation 
\begin{equation}
g(s,z,t)=e^{-t^2/2}\left[{\eta\over 2}-
{\psi'(\eta)\over \psi(\eta)}\right], \qquad \eta=s+t,
\label{trans}
\end{equation}
(prime denotes the derivative with respect to $\eta$) reduces
Eq.~(\ref{eqgszt2}) to the parabolic cylinder equation \cite{Bender}
for the auxiliary function $\psi(\eta)$, namely
$\psi''+(z-1/2-\eta^2/4)\psi=0$. Finally, the generating function 
is found
\begin{equation}
g(s,z,t)=e^{-t^2/2}\left[{t+s\over 2}-
{D'_{z-1}(s+t)+D'_{z-1}(-s-t)\over D_{z-1}(s+t)+D_{z-1}(-s-t)}\right]
\label{gszt2}
\end{equation}
with $D_{z-1}$ the parabolic cylinder function of order $z-1$.  
It can be verified that $N(t)=g(0,1,t)=te^{-t^2/2}$.  
Two important cases are the Laplace
transform of the length distribution function \cite{Sekimoto}
\begin{equation}
g(s,t)=g(s,z=1,t)=e^{-t^2/2}\left[s+t-
{se^{t^2/2+st}\over 1+s\int_0^t d\tau e^{\tau^2/2+s\tau} }\right], 
\label{gst2}
\end{equation}
and the generating function $g(z,t)=\sum_{n\geq 1} z^n g_n(t)$
\begin{equation}
g(z,t)=g(s=0,z,t)=e^{-t^2/2}\left[{t\over 2}-{D'_{z-1}(t)+D'_{z-1}(-t)
\over D_{z-1}(t)+D_{z-1}(-t)}\right].
\label{gzt1}
\end{equation}

We are unable to perform the inverse Laplace transform. However, by a
direct solution of equation (\ref{eqgnxt}), it is still possible to
obtain several quantities.  For example, the one-seed island
distribution obeys
\begin{equation}
\left({\partial\over\partial t}+{\partial\over\partial x}\right)g_1(x,t)
=-2tg_1(x,t)+e^{-t^2/2}\delta(x).  
\label{eg1xt}
\end{equation}
Solving this partial differential equation gives
\begin{equation}
g_1(x,t)=e^{-t^2}e^{(x-t)^2/2}\qquad 0<x<t, 
\label{g1xt}
\end{equation}
and $g(x,t)=0$ outside the space interval $[0, t]$. The above
distribution is strongly peaked ate $x=t$.  A numerical solution of
Eq.~(\ref{eqgxt}) shows that the total gap distribution, $g(x,t)$, is
also sharply peaked at this point \cite{Sekimoto}. In fact, we expect
that one-seed islands dominate for sizes approaching $t$ from
below. Following the previous section findings, we also expect that
the $n$-seed densities are discontinuous on the rays $x_n(t)=nt$.
However, such discontinuous behavior might be hardly visible for large
$n$.

The density of one-seed island equals $g_1(t)=e^{-t^2}\int_0^t d\tau\,  
e^{\tau^2/2}$.  Evaluating the limit $t\to\infty$ gives
$g_1(t)\simeq t^{-1}e^{-t^2/2}$. Unlike the simultaneous case, where
the fraction of one-seed islands decayed exponentially, here we find
that the quantity $p_1(t)\equiv g_1(t)/N(t)\simeq t^{-2}$ decays
significantly slower in an algebraic fashion. This power-law behavior
is actually a general one. Writing $g_n(t)=e^{-t^2/2}\tilde
g_n(t)$ and integrating equation (\ref{eqgnxt}) with respect to $x$,
we find
\begin{equation}
{d\tilde g_n(t)\over dt}= \delta_{n,1}-t\tilde g_n(t)+\sum_{m=1}^{n-1} 
\tilde g_m(t)\tilde g_{n-m}(t).
\label{tildegnt}
\end{equation} 
Instead of solving the above equation generally, we obtain the leading
asymptotic behavior in the limits of small and large $t$.  In the long
time limit, the left hand side is negligible and can be safely
discarded. In the limit of small time, the loss term $-t\tilde g_n(t)$
is unimportant.  Solving the resulting approximate equations gives
\begin{equation}
p_n(t)={g_n(t)\over N(t)}\simeq\cases{
a_n t^{2(n-1)},  &$t\to 0;$\cr
b_n t^{-2n},     &$t\to\infty;$\cr}
\label{pnt2}
\end{equation}
with the prefactors $a_n=2^{2n}(2^{2n}-1)B_n/(2n)!$ ($B_n$ are the
Bernoulli numbers) and $b_n=(2n)!/[2(2n-1)n!^2]$.  While the early
time behavior resembles the simultaneous case, the long time behavior
is algebraic rather than exponential. Note also the nontrivial
dependence of both prefactors, $a_n$ and $b_n$, on the number of seeds
$n$. 

\section{Accelerated Nucleation-and-Growth}

Most of the studies of the nucleation-and-growth processes assume that
the growth velocity is constant (see e.g. \cite{Sekimoto} and
references therein).  In this Section we demonstrate that a particular
generalized KAJM model with {\it accelerated} growth, namely with
velocity linear in the island size, can be treated analytically in one
dimension.  We emphasize that the linear dependence of the growth
velocity on island size naturally appears in several problems.  One
application, mentioned in the Introduction, concerns accelerated
random sequential adsorption (RSA) on a line with precursor layer
diffusion.  Indeed, in accelerated RSA particles are deposited onto
the substrate and they occupy empty sites. Particles that are deposited on
occupied sites (extrinsic precursor state) can lose enough kinetic
energy, such that they do not desorb back to the gas phase. Instead,
they diffuse on top of occupied islands until they encounter an empty
site on the island boundary where they are deposited irreversibly
\cite{king,Filipe}.  If the process is adsorption-limited, {\it i.e.},
the diffusion time scale is small compare to the adsorption time
scale, islands on the 1D substrate grow with rate proportional to
the island length.  Another possible application is related to
biological growth where seeds are the source of the new phase.  Therefore, 
in 1D one-seed islands grow with rate 1, two-seed islands grow with
rate 2, {\it etc.}  The definitions of both accelerated RSA and the
biological model are appealingly simple, and they exhibit similar
behavior to that of the KAJM model with linear growth rate. The latter
model has the advantage that it is more amenable to analytical treatment. 

Thus, we consider the KAJM model with an accelerated growth mechanism,
namely with growth rate proportional to the length of a domain,
or better equal to $1+x$ to ensure a finite growth velocity for
initial size-less islands.  As the length of an isolated domain grows
exponentially in time, it is expected that for this accelerated KAJM
space is covered in {\it finite} time $t_c$.  Below, we will find
exact values for some interesting quantities, including $t_c$ and the
gap distribution.  The rate equation for $f(x,t)$ is a generalization
of equation (\ref{eqfxt}),
\begin{equation}
{\partial f(x,t)\over \partial t}= [1+\langle x(t)\rangle]{\partial
f(x,t)\over \partial x}-\gamma(t) xf(x,t)+2\gamma(t) \int_x^\infty dy\, f(y,t).
\label{eqfxt1}
\end{equation}
The growth term reflects the fact that the growth velocity of islands
is linear in their length. The linear dependence is crucial in writing
the above equation. A gap between two islands of sizes $x_1$ and $x_2$
shrinks with rate $1+(x_1+x_2)/2$, and we can use the equality
$\langle x_1+x_2\rangle/2=\langle x\rangle$.  It is possible to write
down the rate equation for the island distribution function.  However,
the analysis of that equation is very cumbersome.  Therefore in the
following we limit ourselves to the gap distribution which can be
examined in depth for both the instantaneous and continuous 
nucleation.  We also present an approximate treatment of generalized
KAJM models where the growth rate equals an arbitrary power of the
island size.

\subsection{Instantaneous Nucleation}

We consider first the case of instantaneous nucleation. To solve
Eq.~(\ref{eqfxt1}) with $\gamma(t)=\delta(t)$, it is useful to introduce 
a modified time variable, $T(t)$, defined by $T(t)=\int_0^t
dt'[1+\langle x(t')\rangle]$.  In terms of this variable, 
Eq.~(\ref{eqfxt1}) simplifies to 
\begin{equation}
\left({\partial\over \partial T}
-{\partial\over \partial x}\right)f(x,t)=0,
\label{eqfxt2}
\end{equation}
Similar to the usual KAJM growth, the gap distribution is readily found
for arbitrary initial conditions, $f(x,t=0)=f_0(x)$, to yield
$f(x,t)=f_0(x+T)$.  In particular, let us assume that size-less islands
were initially randomly distributed with unit density,
$f_0(x)=e^{-x}$.  Then the gap distribution reads 
\begin{equation}
f(x,t)=e^{-x-T}. 
\label{fxt3}
\end{equation}
Consequently, we get
\begin{equation}
N(T)=S(T)=e^{-T}, \qquad
\langle x(T)\rangle=e^T-1,
\label{NSx}
\end{equation}
The average island length was obtained using $\langle x(T)\rangle=
\big(1-S(T)\big)/N(T)$.  for the number density, the uncovered fraction, and
the average island length, respectively.  To obtain the explicit time
dependence, it is necessary to solve $dT/dt=1+\langle
x(T)\rangle=e^{T}$, which is integrated to yield $e^{-T}=1-t$.  This
allows us to determine the time of complete coverage, $t_c=1$.
Reexpressing the exact results in terms of the physical time, we
arrive at
\begin{equation}
f(x,t)=(1-t)e^{-x}, \qquad
N(t)=S(t)=1-t, \qquad
\langle x(t)\rangle={t\over 1-t}.
\label{fxt4}
\end{equation}
Interestingly, the uncovered fraction and the number density are
equal, as for the KAJM model.  The covered fraction exhibits a linear
growth, $1-S(t)=t$.  This result agrees with the lattice adsorption
process of Ref.~\cite{Filipe}.  Identical behavior is found for the
biological growth process where $n$-seed islands grow with rate $n$.
For simultaneous nucleation, the number of seeds remains constant and
hence the covered fraction increases with constant rate.  However, the
covered fraction is the {\it only} characteristic which is known
analytically for both of these lattice models.

Return now to the accelerated KAJM model and consider the competition
between the two components of the velocity: the intrinsic part and the
size dependent part.  Suppose that size-less islands grow with
velocity $v_0$, {\it i.e.}, the growth velocity of an $x$-island
equals $v_0+x$. Then, the rate equation reads
\begin{equation}
\left({\partial\over \partial t}
-[v_0+\langle x(t)\rangle]{\partial\over \partial x}\right)f(x,t)=0.
\label{eqfxt3}
\end{equation}
The above treatment holds with the modified time variable, $T=\int_0^t
[v_0+\langle x(t') \rangle]dt'$. Following the steps that led to 
equation (\ref{fxt4}) gives 
\begin{equation}
f(x,t)={1-v_0 e^{(1-v_0)t}\over 1-v_0}\, e^{-x}. 
\label{fxt5}
\end{equation}
Additionally, the number density is given by $N(t)=S(t)=(1-v_0
e^{(1-v_0)t})/(1-v_0)$, and the average length is $\langle
x(t)\rangle=v_0(1-e^{(1-v_0)t})/(v_0e^{(1-v_0)t}-1)$.  By taking the
limit $v_0\to 1$, one can verify that the previous results are
recovered.  The critical time is thus $t_c=\ln v_0/(v_0-1)$.  Both
limiting cases exhibit logarithmic behavior
\begin{equation}
t_c\simeq\cases{
\ln (1/v_0),  &$v_0\to 0$;\cr
\ln v_0/v_0,  &$v_0\to \infty$.\cr}
\label{tc1}
\end{equation}
The limiting case $v_0\to 0$ reduces to the KAJM growth where the
coverage time is infinite. When $t\to t_c$, the gap distribution vanishes
according to 
\begin{equation}
f(x,t)\simeq (t_c-t)e^{-x}.
\label{fxtc}
\end{equation} 
It is seen that near-critical behavior such as $S(t)\simeq (t_c-t)$ is
independent of the relative magnitudes of the two components of the
growth velocity.

\subsection{Continuous Nucleation}

We turn now to the continuous nucleation case, $\gamma>0$.  We will
assume that the system is initially empty; therefore, the initial conditions
are given by Eq.~(\ref{data}).  Similar to the previous section, 
we seek for a solution to Eq.~(\ref{eqfxt1}) of the form
\begin{equation}
f(x,t)=\phi(t)e^{-\gamma x t}.
\label{ansatz}
\end{equation}
The choice of the exponential factor allows us to cancel the
$x$-dependent term in the right-hand side of equation
(\ref{eqfxt1}). The rate equation reduces to an ordinary differential
equation,
\begin{equation}
{d\phi(t)\over dt}=2\phi(t)/t-\gamma\phi(t)t[1+\langle x(t)\rangle].
\label{eqphit}
\end{equation}
Expressing $\langle x(t)\rangle$ via $\phi(t)$ gives
$\langle x(t)\rangle=(1-\int_0^\infty dx\,xf(x,t))/\int_0^\infty dx\,f(x,t)
= [1-\phi(t)/(\gamma t)^2]/[\phi(t)/\gamma t]$.
Substituting $\langle x(t)\rangle$ in equation (\ref{eqphit}) and solving the
resulting {\it closed} differential equation for $\phi(t)$, we obtain 
\begin{equation}
f(x,t)=(\gamma t)^2\,e^{-\gamma x t+t-\gamma t^2/2}
\left[1-\int_0^t d\tau\,e^{-\tau+\gamma\tau^2/2}\right].
\label{fxt6}
\end{equation}
This exact solution agrees with the initial condition of equation
(\ref{data}). 

The instant of complete coverage is determined from the positivity
condition of the gap-size distribution function. Thus, $t_c$ is found
from the implicit relation
\begin{equation}
1=\int_0^{t_c} d\tau\,e^{-\tau+\gamma\tau^2/2}.
\label{cond}
\end{equation}
In the limiting situations of small and large birth rate one has
\begin{equation}
t_c\sim \cases{\ln(1/\gamma),              & $\gamma \ll 1;$\cr
               \sqrt{\ln \gamma/\gamma}, & $\gamma \gg 1$.\cr}
\label{tc2}
\end{equation}
Comparing to Eq.~(\ref{tc1}), it is seen that the parameter $\gamma$ plays a
similar role to $v_0$.  Furthermore, in the vicinity of  complete
coverage, $1-t/t_c\ll 1$, the gap distribution function approaches 
\begin{equation}
f(x,t)\simeq (\gamma t_c)^2(t_c-t)e^{-\gamma t_c x},
\label{fxtc1}
\end{equation}
which is surprisingly similar to equation (\ref{fxtc}).  The uncovered
fraction is simply $S(t)=t_c-t$ as for the case of simultaneous
nucleation.  Hence, the near-critical behavior is robust, {\it i.e.}
the details of the nucleation are not important.

\subsection{Generalized Accelerated Nucleation}

We turn now to the general accelerated nucleation-and-growth model with 
growth rate proportional to a power of the island length.  
Let us assume that an $x$-island grows with rate $(1+x)^\alpha$.  Constant
and linear growth rates correspond to $\alpha=0$ and $\alpha=1$, 
respectively.

Consider the case of simultaneous nucleation.  We first write an
{\it exact} rate equation satisfied by uncovered fraction:
\begin{equation}
{d S(t)\over dt}=-N(t)\langle (1+x)^\alpha\rangle.
\label{stg}
\end{equation}
Remember that the previously investigated extreme cases of $\alpha=0$
and $\alpha=1$, yield $S(t)=N(t)$ for simultaneous nucleation (the
initial density of seeds is set equal to one).  Physically, it means
that during the evolution process, the average size of inter-island
gaps does not change.  We now {\it assume} this feature for
generalized nucleation-and-growth models as well.  To proceed, we need
to know the $\alpha^{\rm th}$ moment $\langle (1+x)^\alpha\rangle$.
We certainly know the first moment $\langle
x(t)\rangle=\big(1-S(t)\big)/N(t)$.  For $\alpha>0$, we shall estimate
$\langle (1+x)^\alpha\rangle$ by $[1+\langle x\rangle]^\alpha$ which
is exact only for $\alpha=0$ and $\alpha=1$. Inserting $S(t)=N(t)$ and
$\langle (1+x)^\alpha\rangle\approx [1+(1-S)/N]^\alpha=S^{-\alpha}$
into Eq.~(\ref{stg}) we arrive at the {\it approximate} coverage rate
equation

\begin{equation}
{d S\over dt}=-S^{1-\alpha},
\label{astg}
\end{equation}
which is solved to yield
\begin{equation}
S(t)=(1-\alpha t)^{1/\alpha}. 
\label{ast}
\end{equation} 
Although there is no reason to expect that Eq.~(\ref{ast}) provides an
exact quantitative description for $\alpha$ other than 0 or 1, we do
expect that complete coverage is still reached in a finite time and that
the near-critical behavior is described by the exponent $1/\alpha$.

\section{Summary} 

We investigated analytically the nucleation-and-growth process on
one-dimensional substrates.  We examined both constant and size
dependent growth mechanisms. In the case of simultaneous nucleation
the gap distribution has an infinite set of progressively weaker
discontinuous spatial derivatives.  We introduced the joint
island-number density and showed that while exponential decay
characterizes the fraction of islands containing a fixed number of
seeds for simultaneous nucleation, a much slower power-law decay
occurs for continuous nucleation. An accelerated growth mechanism was
shown to give rise to covering in a finite time, and the near-critical
behavior of the system is insensitive to most details of the growth
process.

\vskip 0.2in

E.~B.~ was supported in part by NSF under Award Number 92-08527 and by
the MRSEC Program of the National Science Foundation under Award
Number DMR-9400379. P.~L.~K. was supported in part by a grant from NSF.

\end{document}